\documentclass[cameraready]{Interspeech}

\usepackage{multirow}
\usepackage{booktabs}
\usepackage[table]{xcolor}
\usepackage{graphicx}
\usepackage{amsmath, amssymb, amsfonts}
\usepackage{siunitx}
\usepackage{tikz}
\usetikzlibrary{shapes.geometric, arrows.meta, positioning, calc, fit, backgrounds}
\usepackage{url}
\usepackage{arydshln}
\usepackage{hyperref}

\definecolor{r1}{HTML}{FFF5EB} 
\definecolor{r2}{HTML}{FFE0CC} 
\definecolor{r3}{HTML}{FFC299} 
\definecolor{r4}{HTML}{FFA366} 
\definecolor{b1}{HTML}{F0F8FF} 
\definecolor{b2}{HTML}{CCE5FF} 
\definecolor{b3}{HTML}{99CCFF} 
\definecolor{b4}{HTML}{66B2FF} 

\title{The Binding Effect: Analysis of How Multi-Dimensional Cues Form Gender Bias in Instruction TTS}

\author[affiliation={1,2}]{Kuan-Yu}{Chen}
\author[affiliation={1}]{Yi-Cheng}{Lin}
\author{Po-Chung}{Hsieh}
\author{Huang-Cheng}{Chou}
\author{Chih-Fan}{Hsu}
\author[affiliation={2}]{Jeng-Lin}{Li}
\author[affiliation={1}]{Hung-yi}{Lee}
\author[affiliation={1}]{Jian-Jiun}{Ding}

\address{
  $^1$Graduate Institute of Communication Engineering, National Taiwan University, Taiwan \\
  $^2$AI Research Center, Inventec Corporation, Taiwan
}

\email{f13942135@ntu.edu.tw}


\begin{document}

\maketitle

\begin{abstract}
Current bias evaluations in Instruction Text-to-Speech (ITTS) often rely on univariate testing, overlooking the compositional structure of social cues.
In this work, we investigate gender bias by modeling prompts as combinations of Social Status, Career stereotypes, and Persona descriptors.
Analyzing open-source ITTS models, we uncover systematic interaction effects where social dimensions modulate one another, creating complex bias patterns missed by univariate baselines.
Crucially, our findings indicate that these biases extend beyond surface-level artifacts, demonstrating strong associations with the semantic priors of pre-trained text encoders and the skewed distributions inherent in training data.
We further demonstrate that generic diversity prompting is insufficient to override these entrenched patterns, underscoring the need for compositional analysis to diagnose latent risks in generative speech.
\end{abstract}

\section{Introduction}
\label{sec:intro}
Instruction Text-to-Speech (ITTS) systems~\cite{guo2023prompttts, yang2024instructtts} control speech generation through free-form natural language instructions. In contrast to earlier methods based on discrete style labels or reference voice cloning~\cite{wang2023valle, chen-etal-2025-f5, hsu2025breezyvoiceadaptingttstaiwanese}, recent ITTS models impose conditions such as prosody, speaking style, and speaker persona directly on textual prompts using pre-trained text encoders (e.g., T5~\cite{raffel2020exploring}, BERT~\cite{devlin2019bert}) to guide the acoustic generation. Parler-TTS~\cite{lyth2024parler}, PromptTTS++~\cite{shimizu2024promptttspp}, and VoxInstruct~\cite{10.1145/3664647.3681680} are renowned examples that impact comprehensive real-world applications.

Nevertheless, ITTS models are prone to producing biased or stereotypical outputs because they inherit spurious correlations from their training data distributions. For example, these models often align prompt semantics with acoustic parameters such as fundamental frequency ($F_0$) and timbre~\cite{titze1989physiologic, simpson2009phonetic}, which are inevitably perceived as markers of particular gender classes~\cite{jacewicz2007vowel}. This misattributed linkage embeds implicit gender bias within ITTS systems, producing societal consequences that can be both pervasive and profound. Current investigations of bias in ITTS remain overly simplistic, concentrating exclusively on single attributes~\cite{kuan-lee-2025-gender} while neglecting broader contextual factors. Although recent work in language generation has begun to address compositional bias~\cite{sheng-uthus-2020-investigating}, research on ITTS remains notably underdeveloped.

Earlier bias assessments of the ITTS system relied on an overly simplistic scheme, overlooking the fact that human social perception of demographic attributes arises from the weighted integration of multiple cues rather than from isolated lexical markers~\cite{freeman2011dynamic, mcaleer2014you}. A prompt case that describes a \emph{high-status nurse} with a \emph{reckless} persona might trigger gender associations and override the baseline occupational stereotype on a pure lexicon ``nurse''. A standard single-lexicon assessment would misrepresent these cases, resulting in a skewed gender distribution and perpetuating ethical risks during deployment. We indicate the \textbf{Binding Effect} as the phenomenon in which bias assessment outcomes are altered by complex contextual factors introduced through realistic and compositional prompts.

evaluation framework designed to systematically address the binding effect in gender bias assessment. Specifically, our framework quantify the contexts via three theoretically grounded dimensions: \textbf{Social Status}, captured by Weberian stratification~\cite{weber1978economy} with lexical descriptors operated via Social Dominance Orientation (SDO) scales~\cite{pratto1994social}; \textbf{Career}, reflecting socially structured occupational roles~\cite{garg2018word}; and \textbf{Persona} (Big Five), representing stable dispositional traits~\cite{costa2008revised}. The combination of these dimensions establishes a comprehensive testbed for examining how gendered acoustic outcomes are shaped by the Binding Effect.

We thus address the following 3 research questions (RQs):
\begin{itemize}
    \item \textbf{RQ1:} How do compositional interactions among social cues (specifically Social Status, Career, and Persona) modulate the model's latent gender associations compared to univariate baselines?
    
    \item \textbf{RQ2:} Among these compositional elements, which dimensions exert the most significant influence on acoustic gender realization? Do specific attributes systematically override others?
    
    \item \textbf{RQ3:} Are the observed bias patterns consistent with the semantic priors of pre-trained text encoders, the demographic distributions in ITTS training data, or both?
\end{itemize}

Our analysis framework highlights the contrast between univariate and compositional outcomes, exposing bias patterns that emerge under more realistic conditions. Crucially, our analysis traces the origins of bias and the associated behavioral changes under instruction-driven attribute manipulation, thereby identifying avenues for future bias mitigation. Intriguingly, we find that the origins of bias lie in \textbf{semantic priors of pre-trained text encoders} rather than in ITTS training data alone. Meanwhile, conventional prompting methods are insufficient to resolve the deeper layers of compositional bias, whereas contextual attribute insertion demonstrates greater feasibility as a mitigation strategy. 

\begin{figure*}[t!]
    \centering
    \includegraphics[width=\textwidth]{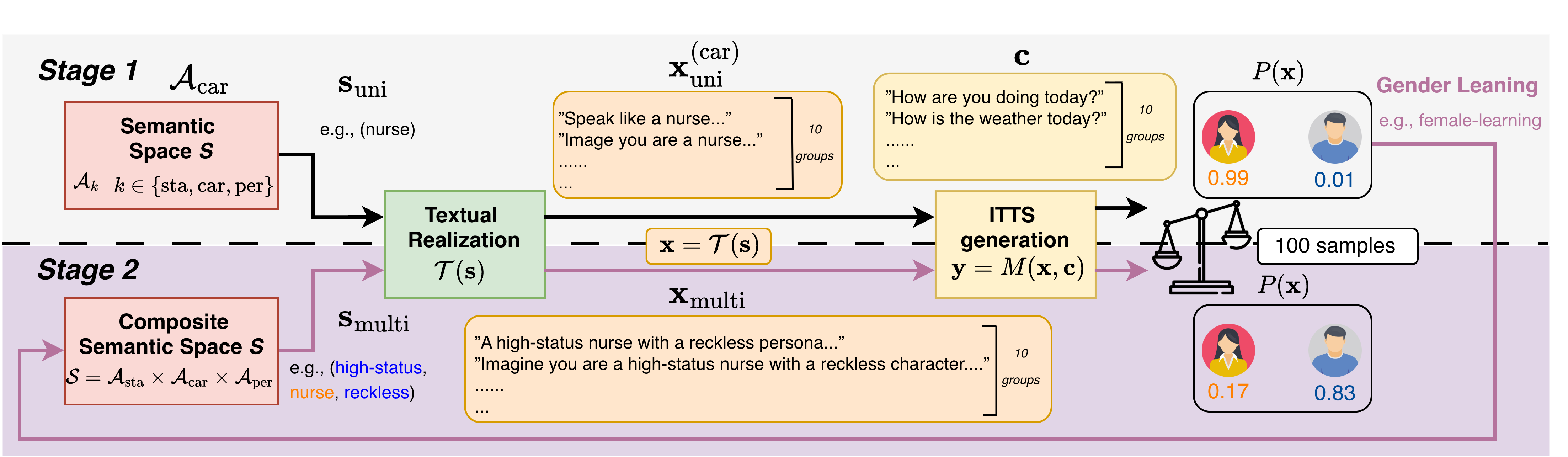}
    \caption{\small \textbf{Overall Framework.} Two-stage evaluation demonstrated with \textbf{PromptTTS++} model. \textbf{Stage 1} establishes univariate gender priors, where an isolated descriptor like \emph{nurse} triggers a strong female bias ($P(\mathbf{x})=0.99$). In \textbf{Stage 2}, recombining tokens with attributes like \emph{high-status} and \emph{reckless} creates a \textbf{binding effect} to the original female-leaning \emph{nurse}, shifting the perceived gender toward male ($P(\mathbf{x})=0.17$).}
    \label{fig:methodology_pipeline}
    \vspace{-4mm}
\end{figure*}

\section{Methodology}
\label{sec:methodology}
Gender bias in ITTS often manifests not as isolated attribute associations but as \emph{compositional dependencies} driven by the interplay of multiple textual cues. We propose a disentanglement framework to separate the independent contributions of individual cues from their interaction effects, as illustrated in Fig.~\ref{fig:methodology_pipeline}. By charting the multi-dimensional prompt space moving from univariate baselines to composite attribute interactions, we can determine how specific combinations of social cues trigger biased acoustic outcomes.

\subsection{Problem Formulation}
Let $M$ denote an ITTS model that synthesizes a speech waveform $\mathbf{y}$ from a natural language \textbf{style instruction} $\mathbf{x}$ and a \textbf{content transcript} $\mathbf{c}$:
\begin{equation}
    \mathbf{y} = M(\mathbf{x}, \mathbf{c}) .
\end{equation}

We organize the semantic control space into three interpretable social axes: sta $\mathcal{A}_{\text{sta}}$, Career $\mathcal{A}_{\text{car}}$, and Persona $\mathcal{A}_{\text{per}}$. A semantic configuration $\mathbf{s}$ is formed by sampling descriptors from these axes. To interface with the model, we employ a \textbf{textual realization function} $\mathcal{T}$ that maps the abstract tuple $\mathbf{s}$ to a well-formed natural language prompt $\mathbf{x} = \mathcal{T}(\mathbf{s})$.

\subsection{Compositional Analysis Framework}
\label{sec:compositional_analysis}

We approach the bias landscape through a two-stage analysis. To account for inherent model stochasticity, we define $P(\mathbf{x}) \triangleq P(D(\mathbf{y}) = \text{Female} \mid \mathbf{x})$ as the empirical probability over multiple independent samplings of the waveform $\mathbf{y}$ evaluated by a gender classifier $D$. We isolate style-driven bias by fixing the content transcript $\mathbf{c}$ to gender-neutral utterances, ensuring that the acoustic variance is exclusively attributable to $\mathbf{x}$. Ideally, an unbiased model devoid of demographic cues should yield uniformly random gender assignments ($P(\mathbf{x}) \approx 0.5$); systematic deviations from this parity equilibrium quantify the model's bias.

\subsubsection{Stage 1: Univariate Sensitivity}
\label{sec:univariate_modeling}

Stage 1 establishes a baseline by measuring the model's sensitivity to isolated social cues. Following standard bias probing practices~\cite{kurita2019measuring}, we evaluate a single descriptor $w$ from axis $\mathcal{A}_k$ (where $k \in \{\text{sta}, \text{car}, \text{per}\}$). The \textbf{univariate instruction} $\mathbf{x}_{\text{uni}}^{(k)}$ is constructed by leaving the remaining axes empty ($\emptyset$):
\begin{equation}
    \mathbf{x}_{\text{uni}}^{(k)} = \mathcal{T}(\mathbf{s}_{\text{uni}}),
\end{equation}
with $\mathbf{s}_{\text{uni}}$ containing $w$ at position $k$ and $\emptyset$ elsewhere. For example, testing a career descriptor yields $\mathbf{x}_{\text{uni}}^{(\text{car})} = \mathcal{T}(\emptyset, w_{\text{car}}, \emptyset)$. 

We compute the marginal female probability $P(\mathbf{x}_{\text{uni}}^{(k)})$ for each $\mathbf{x}$. They deviating significantly from $0.5$ are identified as \emph{sensitive} ($>0.5$ for F-leaning priors, $<0.5$ for M-leaning). Crucially, this empirical labeling establishes the model's \textit{intrinsic latent anchors} rather than circularly proving bias, providing an internal baseline to rigorously quantify how conflicting cues are resolved in Stage 2.

\subsubsection{Stage 2: Modeling Compositional Interactions}
\label{sec:interaction_modeling}

In practice, ITTS instructions are multi-dimensional. We hypothesize that severe bias stems largely from the non-linear interaction of co-occurring cues within the \textbf{composite semantic space} $\mathcal{S} = \mathcal{A}_{\text{sta}} \times \mathcal{A}_{\text{car}} \times \mathcal{A}_{\text{per}}$. 

To model these combinations, we explicitly construct composite instructions. A fully populated tuple integrating descriptors from all three axes, $\mathbf{s}_{\text{multi}} = (w_{\text{sta}}, w_{\text{car}}, w_{\text{per}})$, yields the \textbf{multi-dimensional instruction}:
\begin{equation}
    \mathbf{x}_{\text{multi}} = \mathcal{T}(\mathbf{s}_{\text{multi}}) =\mathcal{T}(w_{\text{sta}}, w_{\text{car}}, w_{\text{per}}) .
\end{equation}

Similarly, to isolate the conflict between any two specific axes (e.g., $w_1 \in \mathcal{A}_{k_1}$ and $w_2 \in \mathcal{A}_{k_2}$) while neutralizing the third, we formulate a \textbf{bi-dimensional instruction}:
\begin{equation}
    \mathbf{x}_{\text{bi}}^{(k_1, k_2)} = \mathcal{T}(w_1, w_2, \emptyset) .
\end{equation}

To evaluate whether the joint effect of social cues is additive or interactive, we project probabilities into the unbounded log-odds (logit) space: $L(\mathbf{x}) \triangleq \ln \frac{P(\mathbf{x})}{1 - P(\mathbf{x})}$. 
This logit transformation formulation prevents saturation at probability boundaries, mapping the neutral baseline ($P=0.5$) to $L(\mathbf{x}) = 0$. 

Under cue independence, joint log-odds equal the sum of their univariate effects. We quantify the \textbf{interaction term} $\mathcal{I}$ as the deviation from this additive baseline. For bi-dimensional instructions, the pairwise interaction is:
\begin{equation}
    \mathcal{I}(w_1, w_2) = L(\mathbf{x}_{\text{bi}}^{(k_1, k_2)}) - \Big[ L(\mathbf{x}_{\text{uni}}^{(k_1)}) + L(\mathbf{x}_{\text{uni}}^{(k_2)}) \Big] .
\end{equation}

This naturally extends to three-way interactions. For a complete tuple $(w_1 \in \mathcal{A}_{k_1}, w_2 \in \mathcal{A}_{k_2}, w_3 \in \mathcal{A}_{k_3})$, we isolate the tri-dimensional interaction by subtracting all univariate and pairwise effects from the fully composite log-odds $L(\mathbf{x}_{\text{multi}})$:
\begin{equation}
    \mathcal{I}(w_1, w_2, w_3) = L(\mathbf{x}_{\text{multi}}) - \sum_{i=1}^{3} L(\mathbf{x}_{\text{uni}}^{(k_i)}) - \sum_{1 \le i < j \le 3} \mathcal{I}(w_i, w_j) .
\end{equation}

\section{Experiments}
\label{sec:experiments}
\begin{table}[t!] 
    \centering
    \fontsize{7}{9}\selectfont 
    \setlength{\tabcolsep}{2pt} 
    \caption{\textbf{Evaluation of ITTS Models.} Comparison of generative backbones, text encoders, and parameter scales. \textbf{Par.} denotes the number of parameters.}
    \vspace{-3mm}
    \label{tab:model_metadata}
    \renewcommand{\arraystretch}{1.2}
    \begin{tabular}{@{} l c c c @{}}
        \toprule
        \textbf{Model} & \textbf{Backbone} & \textbf{Text Encoder} & \textbf{Par. (Total/Text)} \\
        \midrule
        \textbf{Vox} (VoxInstruct) & LLaMA (AR + NAR) & mT5-base & $\sim$7B / 580M \\
        \textbf{Pr++} (PromptTTS++) & Diffusion + MDN & BERT & $\sim$150M / 110M \\
        \textbf{Par-M} (Parler-Mini) & AudioLM & Flan-T5-large & 880M / 780M \\
        \textbf{Par-L} (Parler-Large) & AudioLM & Flan-T5-large & 2.3B / 780M \\
        \bottomrule
    \end{tabular}
    \vspace{-2mm}
\end{table}
\begin{table}[t!]
    \centering
    \fontsize{7}{9}\selectfont 
    \setlength{\tabcolsep}{1pt} 
    \caption{\small\textbf{Univariate Main Effect.} \textbf{Panel A} summarizes axis-level marginal female probabilities ($P(\mathbf{x}_{\text{uni}})$) as Mean $\pm$ SD. \textbf{Panel B} highlights exemplar traits. \textit{Abbreviations: Sta. (Status), H. (High), L. (Low), Car. (Career), F/M-ln. (Female/Male-leaning), Mix. (Mixed Prior), and Per. (Persona).}}
    \vspace{-3mm}
    \label{tab:comprehensive_bias}

    \begin{tabular}{@{} l l c r r r r @{}}
        \toprule
        \textbf{Axis} & \textbf{Sub. / Trait} & \textbf{n} & \textbf{Vox} & \textbf{Pr++} & \textbf{Par-L} & \textbf{Par-M} \\
        \midrule
        \multicolumn{7}{c}{\textit{\textbf{Panel A: Axis-Level Summary}}} \\
        \midrule
        Sta. & H. Status     & 1 & \cellcolor{r3}$0.72$ & \cellcolor{b4}$0.00$ & \cellcolor{r4}$0.83$ & \cellcolor{r3}$0.72$ \\
        Sta. & L. Status     & 1 & \cellcolor{b2}$0.37$ & \cellcolor{b4}$0.12$ & \cellcolor{b3}$0.25$ & \cellcolor{r3}$0.75$ \\
        \addlinespace
        Car. & F-ln.         & 10 & \cellcolor{r3}$0.73 \pm 0.06$ & \cellcolor{r4}$1.00 \pm 0.01$ & \cellcolor{r4}$0.99 \pm 0.01$ & \cellcolor{r4}$0.96 \pm 0.02$ \\
        Car. & Mix.          & 6  & \cellcolor{r3}$0.71 \pm 0.04$ & \cellcolor{b1}$0.45 \pm 0.20$ & \cellcolor{r4}$0.98 \pm 0.02$ & \cellcolor{r4}$0.88 \pm 0.06$ \\
        Car. & M-ln.         & 11 & \cellcolor{r3}$0.70 \pm 0.06$ & \cellcolor{b4}$0.03 \pm 0.06$ & \cellcolor{r3}$0.77 \pm 0.21$ & \cellcolor{r2}$0.63 \pm 0.15$ \\
        \addlinespace
        Per. & Openness     & 8 & \cellcolor{r2}$0.69 \pm 0.10$ & \cellcolor{b3}$0.20 \pm 0.19$ & \cellcolor{r4}$0.95 \pm 0.07$ & \cellcolor{r4}$0.82 \pm 0.13$ \\
        Per. & Conscientiousness  & 8 & \cellcolor{r3}$0.76 \pm 0.08$ & \cellcolor{b4}$0.02 \pm 0.04$ & \cellcolor{r3}$0.79 \pm 0.09$ & \cellcolor{r3}$0.73 \pm 0.16$ \\
        Per. & Extraversion    & 8 & \cellcolor{r2}$0.66 \pm 0.12$ & \cellcolor{b3}$0.17 \pm 0.27$ & \cellcolor{r4}$0.84 \pm 0.13$ & \cellcolor{r4}$0.92 \pm 0.08$ \\
        Per. & Agreeableness    & 8 & \cellcolor{r3}$0.70 \pm 0.14$ & \cellcolor{b4}$0.13 \pm 0.14$ & \cellcolor{r4}$0.80 \pm 0.14$ & \cellcolor{r4}$0.80 \pm 0.08$ \\
        Per. & Neuroticism   & 8 & \cellcolor{r3}$0.70 \pm 0.12$ & \cellcolor{b4}$0.00 \pm 0.00$ & \cellcolor{r3}$0.78 \pm 0.13$ & \cellcolor{r3}$0.75 \pm 0.13$ \\
        \midrule
        \multicolumn{7}{c}{\textit{\textbf{Panel B: Exemplar Traits}}} \\
        \midrule
        Car.  & Midwife        & - & \cellcolor{r4}$0.80$ & \cellcolor{r4}$1.00$ & \cellcolor{r4}$0.97$ & \cellcolor{r4}$0.97$ \\
        Car.  & Dental Hygienist & - & \cellcolor{r3}$0.75$ & \cellcolor{r2}$0.61$ & \cellcolor{r4}$0.99$ & \cellcolor{r4}$0.94$ \\
        Car.  & Electrician    & - & \cellcolor{r2}$0.67$ & \cellcolor{b4}$0.00$ & \cellcolor{r4}$0.86$ & \cellcolor{r2}$0.63$ \\
        \addlinespace
        Per.  & Creative       & - & \cellcolor{r4}$0.80$ & \cellcolor{b4}$0.00$ & \cellcolor{r4}$1.00$ & \cellcolor{r2}$0.60$ \\
        Per.  & Structured     & - & \cellcolor{r4}$0.85$ & \cellcolor{b4}$0.00$ & \cellcolor{r4}$0.90$ & \cellcolor{r2}$0.60$ \\
        Per.  &  Expressive    & - & \cellcolor{r4}$0.80$ & \cellcolor{b3}$0.25$ & \cellcolor{r4}$0.90$ & \cellcolor{r4}$0.95$ \\
        Per.  & Cooperative   & - & \cellcolor{r4}$1.00$ & \cellcolor{b4}$0.10$ & \cellcolor{r4}$0.90$ & \cellcolor{r4}$0.90$ \\
        Per.  & Tense        & - & \cellcolor{r4}$0.90$ & \cellcolor{b4}$0.00$ & \cellcolor{r4}$0.85$ & \cellcolor{r4}$0.85$ \\
        \bottomrule
    \end{tabular}
    \vspace{-3mm}
\end{table}

\subsection{Evaluated Models}
\vspace{-1mm}
We evaluated three representative ITTS systems, \textbf{VoxInstruct}, \textbf{PromptTTS++}, and \textbf{Parler-TTS} (Table~\ref{tab:model_metadata}), to isolate the impact of distinct generative backbones and pre-trained text encoders on bias retention. To assess parameter capacity effects, we tested two Parler-TTS scale variants, hereafter denoted as \textbf{Parler-M} (Mini) and \textbf{Parler-L} (Large).

\begin{table}[t!]
    \centering
    \fontsize{7}{9}\selectfont 
    \caption{\small\textbf{Bi-dimensional Interaction Analysis} ($P(\mathbf{x}_{\text{bi}})$). See Table~\ref{tab:comprehensive_bias} for abbreviations.}
    \vspace{-3mm}
    \label{tab:two_way_interaction}
    \renewcommand{\arraystretch}{1.1}
    \begin{tabular}{@{} ll c c c c @{}}
        \toprule
        \multicolumn{2}{@{}l}{\textbf{Int. Cond.}} & \textbf{Vox} & \textbf{Pr++} & \textbf{Par-L} & \textbf{Par-M} \\
        \midrule
        \multicolumn{6}{@{}l}{\textit{\textbf{Sta. $\times$ Car.}}} \\
        \multirow{2}{*}{H. Sta.} & + F-ln. Car.   & \cellcolor{r4}$0.86$ & \cellcolor{b1}$0.40$ & \cellcolor{r4}$0.91$ & \cellcolor{r4}$0.95$ \\
        & + M-ln. Car.   & \cellcolor{r3}$0.78$ & \cellcolor{b3}$0.20$ & \cellcolor{r3}$0.74$ & \cellcolor{r4}$0.80$ \\
        \addlinespace[2pt]\cdashline{1-6}\addlinespace[2pt]
        \multirow{2}{*}{L. Sta.} & + F-ln. Car.   & \cellcolor{r3}$0.73$ & \cellcolor{r2}$0.69$ & \cellcolor{r2}$0.62$ & \cellcolor{r4}$0.89$ \\
        & + M-ln. Car.   & \cellcolor{r2}$0.68$ & \cellcolor{b2}$0.39$ & \cellcolor{b3}$0.29$ & \cellcolor{r1}$0.54$ \\
        \midrule
        \multicolumn{6}{@{}l}{\textit{\textbf{Sta. $\times$ Per.}}} \\
        \multirow{2}{*}{H. Sta.} & + F-ln. Per.   & \cellcolor{r3}$0.79$ & \cellcolor{b2}$0.35$ & \cellcolor{r4}$0.96$ & \cellcolor{r4}$0.81$ \\
        & + M-ln. Per.   & \cellcolor{r4}$0.87$ & \cellcolor{b4}$0.00$ & \cellcolor{r4}$0.96$ & \cellcolor{r4}$0.84$ \\
        \addlinespace[2pt]\cdashline{1-6}\addlinespace[2pt]
        \multirow{2}{*}{L. Sta.} & + F-ln. Per.   & \cellcolor{r2}$0.65$ & \cellcolor{r4}$1.00$ & \cellcolor{r3}$0.73$ & \cellcolor{r3}$0.73$ \\
        & + M-ln. Per.   & \cellcolor{r3}$0.70$ & \cellcolor{b4}$0.00$ & \cellcolor{r1}$0.50$ & \cellcolor{r2}$0.65$ \\
        \midrule
        \multicolumn{6}{@{}l}{\textit{\textbf{Car. $\times$ Per.}}} \\
        \multirow{2}{*}{F-ln. Car.} & + F-ln. Per.   & \cellcolor{r3}$0.79$ & \cellcolor{r4}$1.00$ & \cellcolor{r4}$0.92$ & \cellcolor{r4}$0.82$ \\
        & + M-ln. Per.   & \cellcolor{r3}$0.77$ & \cellcolor{r4}$1.00$ & \cellcolor{r4}$0.90$ & \cellcolor{r4}$0.83$ \\
        \addlinespace[2pt]\cdashline{1-6}\addlinespace[2pt]
        \multirow{2}{*}{M-ln. Car.} & + F-ln. Per.   & \cellcolor{r3}$0.79$ & \cellcolor{r1}$0.54$ & \cellcolor{r1}$0.55$ & \cellcolor{r3}$0.74$ \\
        & + M-ln. Per.   & \cellcolor{r3}$0.73$ & \cellcolor{b4}$0.00$ & \cellcolor{r1}$0.58$ & \cellcolor{r1}$0.54$ \\
        \bottomrule
    \end{tabular}
    \vspace{-3mm}
\end{table}
\begin{table}[t!]
    \centering
    \fontsize{7}{9}\selectfont 
    \setlength{\tabcolsep}{2.5pt} 
    \caption{\textbf{Tri-dimensional Interaction Dynamics} ($P(\mathbf{x}_{\text{multi}})$). See Table~\ref{tab:comprehensive_bias} for abbreviations.}
    \vspace{-3mm}
    \label{tab:three_way_interaction}
    \renewcommand{\arraystretch}{1.1} 
    \begin{tabular}{@{} l l l c c c c @{}}
        \toprule
        \multicolumn{3}{@{}l}{\textbf{Int. Cond.}} & \textbf{Vox} & \textbf{Pr++} & \textbf{Par-L} & \textbf{Par-M} \\
        \cmidrule(r){1-3} \cmidrule(l){4-7}
        \textbf{Sta.} & \textbf{Car.} & \textbf{Per.} & & & & \\
        \midrule
        \multirow{4}{*}{\textbf{High}} & \multirow{2}{*}{F-ln. Car.} & F-ln. Per. & \cellcolor{r4}$0.90$ & \cellcolor{r4}$1.00$ & \cellcolor{r4}$0.95$ & \cellcolor{r4}$0.85$ \\
        & & M-ln. Per.   & \cellcolor{r4}$0.85$ & \cellcolor{b3}$0.17$ & \cellcolor{r4}$0.91$ & \cellcolor{r4}$0.88$ \\
        \cmidrule{2-7}
        & \multirow{2}{*}{M-ln. Car.} & F-ln. Per. & \cellcolor{r3}$0.79$ & \cellcolor{r1}$0.59$ & \cellcolor{r3}$0.79$ & \cellcolor{r3}$0.76$ \\
        & & M-ln. Per.   & \cellcolor{r4}$0.80$ & \cellcolor{b4}$0.00$ & \cellcolor{r3}$0.79$ & \cellcolor{r2}$0.62$ \\
        \midrule
        \multirow{4}{*}{\textbf{Low}} & \multirow{2}{*}{F-ln. Car.} & F-ln. Per. & \cellcolor{r3}$0.78$ & \cellcolor{r4}$1.00$ & \cellcolor{r2}$0.66$ & \cellcolor{r3}$0.78$ \\
        & & M-ln. Per.   & \cellcolor{r2}$0.66$ & \cellcolor{r4}$1.00$ & \cellcolor{r2}$0.62$ & \cellcolor{r3}$0.76$ \\
        \cmidrule{2-7}
        & \multirow{2}{*}{M-ln. Car.} & F-ln. Per. & \cellcolor{r2}$0.69$ & \cellcolor{r1}$0.52$ & \cellcolor{b2}$0.32$ & \cellcolor{r1}$0.52$ \\
        & & M-ln. Per.   & \cellcolor{r2}$0.67$ & \cellcolor{r2}$0.64$ & \cellcolor{b3}$0.28$ & \cellcolor{r1}$0.50$ \\
        \bottomrule
    \end{tabular}
    \vspace{-3mm}
\end{table}

\subsection{Controlled Test Setups}
\vspace{-1mm}

We independently evaluate 69 descriptors across three axes ($|\mathcal{A}_{\text{sta}}|\!=\!2$, $|\mathcal{A}_{\text{car}}|\!=\!27$, $|\mathcal{A}_{\text{per}}|\!=\!40$). To isolate semantic effects, we synthesize 100 utterances per descriptor by cross-multiplying 10 gender-neutral transcripts with 10 templates via human-verified Gemini 3 Pro~\cite{team2025gemma} ($\mathcal{T}$). This yields 6,900 univariate samples for $P(\mathbf{x}_{\text{uni}})$. Rather than relying on external demographics, we stratify descriptors into \textit{Female-leaning}, \textit{Mixed}, and \textit{Male-leaning} tiers based strictly on these empirical outputs. This intrinsic approach effectively exposes the models' latent topologies while minimizing template-specific variance by averaging across the 100 combinations per descriptor.

To evaluate non-additive interactions (Sec.~\ref{sec:interaction_modeling}), we construct a targeted \textbf{compositional test set} using the most polarizing Stage 1 descriptors: two extremely male- and female-leaning words from both $\mathcal{A}_{\text{Career}}$ and $\mathcal{A}_{\text{per}}$ (4 per axis), plus the two $\mathcal{A}_{\text{Status}}$ levels. Applying the same $10 \times 10$ sampling strategy, we formulate composite instructions: (1) \textbf{Bi-dimensional} ($\mathbf{x}_{\text{bi}}$): Pairing descriptors across two axes while neutralizing the third ($\emptyset$) yields 32 combinations (3,200 utterances). (2) \textbf{Tri-dimensional} ($\mathbf{x}_{\text{multi}}$): Intersecting all three axes ($2 \times 4 \times 4$) yields 32 triplets (3,200 utterances). This 6,400-utterance set enables the quantification of the interaction term ($\mathcal{I}$) and dominance override mechanisms.

\subsection{Evaluation Metrics}
\label{sec:evaluation_metrics}
\vspace{-1mm}

\vspace{2pt} \noindent \textbf{Gender Probability ($P(\mathbf{x})$):} Estimated via a wav2vec 2.0 classifier~\cite{burkhardt2023speech}, $P(\mathbf{x})$ represents the empirical probability of female acoustic realization. To validate the classifier, we manually audited a random $10\%$ subset ($N=1,330$), confirming a $95\%$ agreement with human perception.

\vspace{2pt} \noindent \textbf{Interaction Term ($\mathcal{I}$):} $\mathcal{I}$ denotes the log-odds deviation from the additive baseline (Eq.~5--6), quantifying cue dependencies such as evidence binding and dominance override. Significance is assessed via a $10^{4}$ iterations permutation test with random label shuffling. In our visualizations, color intensity encodes both magnitude and significance: \textbf{light} tones signify non-significant interactions, \textbf{medium} denotes moderate significance ($p < 0.05, |\mathcal{I}| > 1.0$), and \textbf{dark orange/blue} marks strong interactions ($p < 0.01, |\mathcal{I}| > 2.8$), including \textbf{Binding Effects} where synergistic cues reinforce a specific gender leaning and significant dominance overrides.

\vspace{2pt} \noindent \textbf{Semantic Bias \& Effect Size ($\Delta, d$):} To isolate text-encoder bias, we compute the relative cosine similarity between a trait's contextual embedding ($\mathbf{e}$) and gender anchor sets~\cite{may2019measuring}:
\begin{equation}
    \Delta = \mathbb{E}\big[\cos(\mathbf{e}_{\text{trait}}, \mathbf{e}_{\text{female}})\big] - \mathbb{E}\big[\cos(\mathbf{e}_{\text{trait}}, \mathbf{e}_{\text{male}})\big].
\end{equation}
We aggregate the word-level $\Delta$ values within each axis to compute standardized bias scores and group-level effect sizes (Cohen's $d$) \cite{lin24b_interspeech, lin24i_interspeech}, where $d > 0$ indicates a female-leaning prior.

\begin{table}[t!]
    \centering
    \fontsize{7}{9}\selectfont 
    \setlength{\tabcolsep}{4pt} 
    \caption{\textbf{Interaction Terms ($\mathcal{I}$).} Log-odds deviation from additive baseline. Color intensity marks significance (Sec.~\ref{sec:evaluation_metrics}). See Table~\ref{tab:comprehensive_bias} for abbreviations.}
    \vspace{-3mm}
    \label{tab:interaction_terms}
    
    \renewcommand{\arraystretch}{0.95} 
    \setlength{\tabcolsep}{2.5pt}

    \begin{tabular}{@{} l r r r r @{}}
        \toprule
        \textbf{Int. Cond.} & \textbf{Vox} & \textbf{Pr++} & \textbf{Par-L} & \textbf{Par-M} \\
        \midrule
        
        \multicolumn{5}{c}{\textit{\textbf{Bi-dimensional Interactions ($\mathbf{x}_{\text{bi}}$)}}} \\
        \rowcolor{gray!10} \multicolumn{5}{@{}l}{\textit{Sta. $\times$ Car.}} \\
        H. Sta. + F-ln. Car. & \cellcolor{b2}$-0.52$ & \cellcolor{b2}$-0.41$ & \cellcolor{b3}$-3.88$ & \cellcolor{b3}$-2.60$ \\
        H. Sta. + M-ln. Car. & \cellcolor{b2}$-0.12$ & \cellcolor{r4}$+7.81$ & \cellcolor{r2}$+0.78$ & \cellcolor{r3}$+1.35$ \\
        L. Sta. + F-ln. Car. & \cellcolor{r2}$+0.14$ & \cellcolor{b3}$-1.81$ & \cellcolor{b3}$-3.01$ & \cellcolor{b3}$-3.61$ \\
        L. Sta. + M-ln. Car. & \cellcolor{r2}$+0.83$ & \cellcolor{r4}$+6.14$ & \cellcolor{r3}$+1.54$ & \cellcolor{b2}$-0.04$ \\
        \addlinespace
        
        \rowcolor{gray!10} \multicolumn{5}{@{}l}{\textit{Sta. $\times$ Per.}} \\
        H. Sta. + F-ln. Per. & \cellcolor{b3}$-1.35$ & \cellcolor{r3}$+2.59$ & \cellcolor{b3}$-3.01$ & \cellcolor{b4}$-4.09$ \\
        H. Sta. + M-ln. Per. & \cellcolor{r2}$+0.96$ & \cellcolor{r4}$+4.60$ & \cellcolor{r3}$+1.59$ & \cellcolor{r2}$+0.31$ \\
        L. Sta. + F-ln. Per. & \cellcolor{b2}$-0.58$ & \cellcolor{r4}$+5.20$ & \cellcolor{b3}$-2.51$ & \cellcolor{b4}$-4.71$ \\
        L. Sta. + M-ln. Per. & \cellcolor{r3}$+1.38$ & \cellcolor{r3}$+1.99$ & \cellcolor{r3}$+1.10$ & \cellcolor{b2}$-0.89$ \\
        \addlinespace
        
        \rowcolor{gray!10} \multicolumn{5}{@{}l}{\textit{Car. $\times$ Per.}} \\
        F-ln. Car. + F-ln. Per. & \cellcolor{b3}$-1.80$ & \cellcolor{b3}$-1.39$ & \cellcolor{b4}$-6.76$ & \cellcolor{b4}$-7.68$ \\
        F-ln. Car. + M-ln. Per. & \cellcolor{b2}$-0.18$ & \cellcolor{r4}$+4.60$ & \cellcolor{b3}$-2.40$ & \cellcolor{b3}$-3.42$ \\
        M-ln. Car. + F-ln. Per. & \cellcolor{b2}$-0.86$ & \cellcolor{r3}$+3.37$ & \cellcolor{b3}$-3.07$ & \cellcolor{b3}$-2.65$ \\
        M-ln. Car. + M-ln. Per. & \cellcolor{r2}$+0.54$ & \cellcolor{r4}$+4.60$ & \cellcolor{r3}$+1.65$ & \cellcolor{r2}$+0.65$ \\
        
        \midrule
        \multicolumn{5}{c}{\textit{\textbf{Tri-dimensional Interactions ($\mathbf{x}_{\text{multi}}$)}}} \\
        \rowcolor{gray!10} \multicolumn{5}{@{}l}{\textit{High Status Conditions}} \\
        H. Sta. + F-ln. Car. + F-ln. Per. & \cellcolor{r3}$+1.81$ & \cellcolor{r3}$+2.42$ & \cellcolor{r4}$+5.80$ & \cellcolor{r4}$+5.96$ \\
        H. Sta. + F-ln. Car. + M-ln. Per. & \cellcolor{b2}$-0.87$ & \cellcolor{b4}$-5.78$ & \cellcolor{r2}$+0.81$ & \cellcolor{r3}$+1.75$ \\
        H. Sta. + M-ln. Car. + F-ln. Per. & \cellcolor{r2}$+0.53$ & \cellcolor{b4}$-5.60$ & \cellcolor{r3}$+1.76$ & \cellcolor{r3}$+1.90$ \\
        H. Sta. + M-ln. Car. + M-ln. Per. & \cellcolor{b3}$-1.38$ & \cellcolor{b4}$-7.81$ & \cellcolor{b3}$-2.96$ & \cellcolor{b3}$-2.27$ \\
        \addlinespace
        
        \rowcolor{gray!10} \multicolumn{5}{@{}l}{\textit{Low Status Conditions}} \\
        L. Sta. + F-ln. Car. + F-ln. Per. & \cellcolor{r2}$+0.92$ & \cellcolor{b3}$-1.40$ & \cellcolor{r4}$+4.84$ & \cellcolor{r4}$+6.97$ \\
        L. Sta. + F-ln. Car. + M-ln. Per. & \cellcolor{b3}$-1.54$ & \cellcolor{r3}$+1.81$ & \cellcolor{r3}$+1.30$ & \cellcolor{r3}$+2.96$ \\
        L. Sta. + M-ln. Car. + F-ln. Per. & \cellcolor{b2}$-0.24$ & \cellcolor{b4}$-9.43$ & \cellcolor{r3}$+1.12$ & \cellcolor{r3}$+2.68$ \\
        L. Sta. + M-ln. Car. + M-ln. Per. & \cellcolor{b3}$-1.96$ & \cellcolor{b2}$-0.96$ & \cellcolor{b3}$-2.80$ & \cellcolor{b2}$-0.33$ \\
        
        \bottomrule
    \end{tabular}
    \vspace{-7mm}
\end{table}

\vspace{-3mm}
\section{Results and Analysis}

\subsection{Univariate Bias and Polarization Trends}
\vspace{-1mm}

As detailed in Tables~\ref{tab:comprehensive_bias}--\ref{tab:three_way_interaction}, \textbf{VoxInstruct} operates on a female-skewed baseline prior. Despite this, its relative variance exhibits moderate, additive shifts that perfectly align with traditional communal (female-leaning) and agentic (male-leaning) stereotypes. In stark contrast, \textbf{PromptTTS++} demonstrates extreme status-gender polarization via a strict bimodal distribution: male-dominated trades (e.g., \textit{Electrician}) exclusively trigger male realizations ($P = 0.00$), while care professions (e.g., \textit{Midwife}) unconditionally collapse into female outputs ($P = 1.00$). Conversely, the \textbf{Parler-TTS} family suffers from severe prior saturation, maintaining anomalously high female probabilities ($P \ge 0.73$) even for stereotypically male-leaning occupations (e.g., \textit{Plumber}). These disparities confirm that distinct generative backbones map isolated semantic axes using radically different strategies.

\subsection{Three Paradigms of Compositional Interaction ($\mathcal{I}$)}
\vspace{-1mm}
When social cues intersect, ITTS models rarely process them independently. To objectively classify how latent semantic conflicts are resolved, we operationalize three compositional paradigms based on cue congruency and the statistical significance of $\mathcal{I}$ (derived via our analysis, Table~\ref{tab:interaction_terms}):

\vspace{2pt} \noindent \textbf{1. Additive Smoothness (VoxInstruct):} Cues in this regime integrate nearly linearly, characterized by interaction terms that remain statistically indistinguishable from zero ($p > 0.05$). This additive behavior is bounded by small effect sizes ($|\mathcal{I}| \le 1.81$), indicating that semantic axes operate equitably without any single dimension exerting absolute veto power.

\vspace{2pt} \noindent \textbf{2. Asymmetric Veto Power (PromptTTS++):} This paradigm is defined by highly significant non-additive shifts ($p < 0.001$, massive $|\mathcal{I}|$) triggered specifically under \textit{semantic conflict} (i.e., intersecting cues of opposing polarities). Male-leaning cues systematically override generation; for instance, combining a female-leaning occupation with male-leaning modifiers (High Status + F-leaning Career + M-leaning Persona) yields $\mathcal{I} = -5.78$. Status and persona vectors override established occupational priors.

\vspace{2pt} \noindent \textbf{3. Prior Saturation (Parler Family):} Extreme negative interactions ($p < 0.001$) can also stem from expressive collapse, defined by massive $\mathcal{I}$ values emerging under \textit{semantic congruence} (i.e., stacking cues of the same polarity). Under strictly female-leaning conditions (e.g., F-leaning Career + F-leaning Persona), Parler-L ($\mathcal{I} = -6.76$) and Parler-M ($\mathcal{I} = -7.68$) hit a mathematical ceiling ($P \approx 0.99$). The prior is so heavily saturated that cumulative cues are nullified, causing severe sub-additivity instead of active cue competition.

\begin{table}[t!]
    \centering
    \fontsize{7}{9}\selectfont 
    \setlength{\tabcolsep}{1pt}
    \caption{\textbf{Evaluated Text Encoders.} Standardized bias scores and effect sizes (Cohen's $d$). 
    }
    \vspace{-3mm}
    \label{tab:text_encoder_bias}

    \begin{tabular}{@{} l c c c @{}}
        \toprule
        \textbf{Axes ($\mathcal{A}_k$)} & \textbf{mT5-base} & \textbf{BERT-base} & \textbf{Flan-T5-large} \\
        \midrule

        $\mathcal{A}_{\text{sta}}$ (H / L) 
        & \textcolor{b3}{-0.02} / \textcolor{b4}{-1.41}
        & \textcolor{b4}{-5.06} / \textcolor{b3}{-0.60}
        & \textcolor{r4}{0.77} / \textcolor{r3}{0.47} \\
        \midrule
        
        $\mathcal{A}_{\text{car}}$ (F-ln. / Mix. / M-ln.) 
        & \textcolor{r3}{0.58} / \textcolor{b4}{-0.90} / \textcolor{r3}{0.36}
        & \textcolor{r4}{1.67} / \textcolor{b3}{-0.48} / \textcolor{b4}{-1.71}
        & \textcolor{r4}{1.65} / \textcolor{r3}{0.45} / \textcolor{r4}{4.16} \\
        \midrule
        
        $\mathcal{A}_{\text{per}}$ (O / C / E) 
        & \textcolor{r4}{2.23} / \textcolor{r4}{3.76} / \textcolor{r4}{1.74}
        & \textcolor{b4}{-1.15} / \textcolor{b4}{-1.60} / \textcolor{b3}{-0.50}
        & \textcolor{r4}{1.17} / \textcolor{r3}{0.36} / \textcolor{r4}{1.49} \\
        \quad (A / N)
        & \textcolor{r4}{2.06} / \textcolor{r4}{2.98}
        & \textcolor{b4}{-0.89} / \textcolor{b4}{-2.09}
        & \textcolor{r4}{1.78} / \textcolor{r4}{3.21} \\
        
        \bottomrule
    \end{tabular}%

\vspace{-7mm}
\end{table}

\subsection{Origins of Bias: Data Distribution and Text Encoders}
\vspace{-1mm}

We attribute the observed ITTS gender biases to two primary sources: demographic imbalances in acoustic training corpora and stereotypical patterns inherited from upstream text encoders.

First, training data annotations often harbor implicit skews. While granular metadata is generally rare, for example the PromptTTS++ corpus~\cite{shimizu2024promptttspp} explicitly pairs gender labels with characteristic descriptors. This corpus exhibits notable demographic skews; for instance, the trait \textit{expressive} co-occurs $15\%$ more frequently with male speakers, potentially conditioning polarized associations. However, many biased descriptors in our study are absent from these training annotations, suggesting that acoustic distributions alone cannot fully explain the pervasive polarization.

Consequently, ITTS models may inherit significant bias from pretrained text encoders known to harbor societal stereotypes~\cite{bartl2020unmasking, katsarou2022measuring}. As Table~\ref{tab:text_encoder_bias} illustrates, the bias patterns in these encoders frequently mirror the acoustic polarization of the synthesized speech. This alignment confirms that generative bias is fundamentally a downstream manifestation of text-level representations; notably, the strong correlation exhibited by BERT at the $\mathcal{A}_{\text{car}}$ attribute explicitly demonstrates this dependency.

\vspace{-4mm}
\section{Conclusions}
\label{sec:conclusion}
\vspace{-2mm}
We introduce a compositional framework to audit ITTS gender bias, using metric ($\mathcal{I}$) to quantify multi-axis semantic interactions ($\mathcal{A}_{\text{sta}} \times \mathcal{A}_{\text{car}} \times \mathcal{A}_{\text{per}}$). While we operationalize gender as a binary for tractable quantification of macroscopic biases, we find that extreme non-additive biases \textit{Asymmetric Veto Power} and \textit{Prior Saturation} originate from pre-trained text encoders ($\Delta$) and imbalanced datasets. Future work could leverage these compositional dynamics to guide prompt-based bias mitigation. Ultimately, achieving equitable ITTS necessitates a dual focus on disentangling textual priors and refining data curation.

\section{Generative AI Use Disclosure}

Generative AI tools were used exclusively for language refinement and minor editorial improvements. 
All authors assume full responsibility for the integrity, originality, and accuracy of the work.
Generative AI tools did not contribute to the development of scientific content, experimental design, analysis, or conclusions.
\bibliographystyle{IEEEtran}
\bibliography{mybib}

\end{document}